\documentclass{tcibook}
\usepackage{fancyhea}
\usepackage{work}
\usepackage{bm}       
\usepackage{graphicx}
\usepackage{hyperref}      

\usepackage{chngcntr}
\counterwithout{figure}{chapter}

\newcommand{\nc}{\newcommand}  

\def\beq{\begin{equation}}
\def\eeq#1{\label{#1}\end{equation}}
\def\eeqn{\end{equation}}

\newenvironment{Eqnarray}%
   {\arraycolsep 0.14em\begin{eqnarray}}{\end{eqnarray}}
\def\beqa{\begin{Eqnarray}}
\def\eeqa#1{\label{#1}\end{Eqnarray}}
\def\eeqan{\end{Eqnarray}}

\nc{\ra}{\rightarrow}  
\nc{\slsh}{\slash\hspace*{-0.22cm}}
\def\Re{{\cal R \mskip-4mu \lower.1ex \hbox{\it e}\,}}
\def\Im{{\cal I \mskip-5mu \lower.1ex \hbox{\it m}\,}}

\nc{\vev}[1]{ \left\langle {#1} \right\rangle }
\nc{\bra}[1]{ \langle {#1} | }
\nc{\ket}[1]{ | {#1} \rangle }
\nc{\fb}{\,{\rm fb}^{-1}}
\nc{\ev}{{\rm eV}}
\nc{\kev}{{\rm keV}}
\nc{\Mev}{{\rm MeV}}
\nc{\gev}{{\rm GeV}}
\nc{\tev}{{\rm TeV}}
\nc{\mev}{{\rm MeV}}

\def\del{\partial}
\def\Dslash{\not{\hbox{\kern-4pt $D$}}}
\def\dslash{\not{\hbox{\kern-2pt $\del$}}}
\def\pslash{\not{\hbox{\kern-2pt $p$}}}
\def\ETmiss{ \not{\hbox{\kern-4pt $E$}}_T }

\def\eff{{\mbox{\scriptsize eff}}}

\def\msb{{\bar{\ssstyle M \kern -1pt S}}}

\begin{document}

\def\bibname{References}

\bibliographystyle{utphys}  

\raggedbottom

\pagenumbering{roman}

\parindent=0pt
\parskip=8pt
\setlength{\evensidemargin}{0pt}
\setlength{\oddsidemargin}{0pt}
\setlength{\marginparsep}{0.0in}
\setlength{\marginparwidth}{0.0in}
\marginparpush=0pt


\pagenumbering{arabic}

\renewcommand{\arraystretch}{1.25}
\addtolength{\arraycolsep}{-3pt}


\newcommand{\lya}{Ly$\alpha$}
\newcommand{\rref}{\par\noindent\hangindent 15 pt}
\renewcommand*\thesection{\arabic{section}}

\chapter*{Facilities for Dark Energy Investigations}

\begin{center}
\begin{large} {\bf Convener: David Weinberg} \end{large}

Deborah Bard,
Kyle Dawson,
Olivier Dor\'e,
Joshua Frieman,
Karl Gebhardt,
Michael Levi,
Jason Rhodes

\end{center}

\section{Overview}

The discovery of cosmic acceleration has inspired ambitious
experimental and observational efforts to understand its origin.  
Many of these take the form of large astronomical surveys,
sometimes using new, special-purpose instrumentation, and
in some cases entirely new facilities.  These efforts are
frequently referred to as ``dark energy experiments,'' which
is the terminology we will adopt in this Report; we will
not try to draw any fine distinctions between ``experiments''
and ``observations.''  The basic thrust of these experiments
is to measure the history of expansion and growth of structure
in the universe with steadily increasing precision over the
broadest achievable range of redshifts.  The hope is that
these measurements may eventually reveal discrepancies with
the predictions of a simple cosmological constant model,
perhaps pointing towards a particular class of dynamical field
or a breakdown of General Relativity (GR) on cosmological scales.
The approach is in many ways analogous to high-energy physics
experiments that pursue increasing measurement precision or
new ranges of energy, hoping to ``break'' the standard model
in a way that points towards deeper physics that underpins it.

Most existing and planned dark energy experiments employ one or
more of the following approaches:
\begin{enumerate}

\item Wide-field imaging, to measure weak gravitational lensing, 
the galaxy content and weak lensing signal of galaxy clusters,
and angular clustering of galaxies in bins of photometrically
estimated redshift.

\item Synoptic observations to discover and monitor Type Ia supernovae,
using them to measure the distance-redshift relation.

\item Wide-field spectroscopy, to map the clustering of galaxies,
quasars, and the \lya\ forest and thereby measure distances and
expansion rates with baryon acoustic oscillations (BAO) and the
history of structure growth with redshift-space distortions (RSD).

\end{enumerate}
These techniques are reviewed in detail by \cite{weinberg13} and
discussed in many of the Snowmass CF5 Reports.
The same wide-field imaging facilities are often well adapted to
both of the first two approaches, which can be interleaved in
a coherent observing strategy.  However, a supernova survey also
requires spectroscopic observations to determine redshifts and
confirm the Type Ia identification.  These spectroscopic observations
are usually carried out from telescopes other than the ones
used for discovery and photometric monitoring, and the spectroscopic
observations may require significantly more observing time
(and larger aperture telescopes) than the discovery and monitoring themselves.
To relieve the stringent demands of obtaining ``real-time'' spectroscopy,
there is increasing focus on the use of photometrically defined 
Type Ia samples with subsequent spectroscopy to obtain host-galaxy
redshifts, though in these cases one still needs spectroscopic
observations of a subset of SNe to estimate contamination rates.

Large cosmological surveys that have completed their observations,
though not necessarily their final analyses, include the
imaging and spectroscopic surveys of the Sloan Digital Sky Survey
(SDSS-I and II, \cite{abazajian09}),
the SDSS-II supernova survey \cite{frieman08,campbell12},
all conducted from the Sloan Foundation 2.5m telescope in New Mexico
\cite{gunn06},
and the Supernova Legacy Survey (SNLS, \cite{sullivan11}) and
CFHTLens weak lensing survey \cite{heymans13}, both conducted
from the Canada-France-Hawaii 3.6m telescope.

In this Report we briefly summarize some of the major dark
energy experiments that are currently in operation or are planned
to start operations in the next decade.  We restrict our discussion
to projects that are U.S.-led or, in some cases, where U.S. scientists
are playing a significant supporting role.  Many of these projects
are described in publicly available documents that run for tens
or hundreds of pages.  Here we have confined our discussion of each 
project to two pages, with selective references to additional
documentation.  These facilities and experiments are referred to in 
many of the other CF5 Reports.  The projects that we summarize are:
\begin{itemize}

\item The Baryon Oscillation Spectroscopic Survey (BOSS), a wide-field
spectroscopic survey on the Sloan 2.5m telescope, now entering its final 
year of observations.

\item The Dark Energy Survey (DES), a wide-field imaging survey and supernova
survey on the Blanco 4m telescope, using the recently commissioned
Dark Energy Camera, commencing survey operations in Fall 2013
and continuing for five years.

\item The extended Baryon Oscillation Spectroscopic Survey (eBOSS), 
a wide-field spectroscopic survey on the Sloan 2.5m telescope extending
to higher redshifts than the BOSS galaxy survey.  The intended start 
is summer 2014, extending for six years.

\item The Subaru Hyper-Suprime Camera (HSC) and Prime Focus Spectrograph
(PFS), wide-field imaging and spectroscopic instruments that will be
used for cosmological surveys from the Subaru 8.4m telescope.
Intended starts are 2014 (HSC) and 2018 for (PFS).

\item The Hobby-Eberly Telescope Dark Energy Experiment (HETDEX), which
will use wide-field spectroscopy on the HET to survey
\lya\ emitters at $1.9 < z < 3.5$.  Intended start is fall 2014,
with a 3-year duration and extension to 5 years if needed.

\item The Dark Energy Spectroscopic Instrument (DESI), a wide-field
spectroscopic instrument that will carry out dark energy surveys 
from the Mayall 4m telescope.  Intended start is 2018, with 5-year
survey duration.

\item The Large Synoptic Survey Telescope (LSST), a new 8.4m
telescope equipped with an enormous wide-field optical camera, 
dedicated to wide-field imaging and synoptic time-domain surveys.
Intended start is 2020, with 10-year duration.

\item Euclid, a 1.2m space telescope dedicated to a large area 
weak lensing survey with optical imaging and a large area spectroscopic
survey with slitless infrared spectroscopy.  Scheduled launch is 2020,
with 6-year duration of the prime mission.

\item The Wide-Field Infrared Survey Telescope (WFIRST), which in its
most recently proposed design would use a 2.4m telescope for wide-field
imaging and slitless spectroscopic surveys and a supernova survey,
in addition to other astronomical surveys.  Intended launch is
early 2020s, with 6-year duration of the prime mission.

\end{itemize}

Before describing these facilities individually, it is worth making
two general points.  First, the data sets constructed for 
``dark energy experiments'' are very rich, supporting a wide range
of other scientific objectives.  Some of these lie within the 
broader scope of Cosmic Frontiers 
--- in particular, precision measurements of galaxy clustering
and weak lensing are powerful probes of neutrino masses, unknown 
relativistic species, inflation physics, and perhaps the nature
of dark matter.  Others lie outside Cosmic Frontiers, such as
galaxy and quasar evolution, the physics of the intergalactic medium,
and the structure and history of the Milky Way and neighboring galaxies.
This broad scientific reach is a primary reason that the astronomical
community has committed resources, expertise, and effort to these
experiments.  LSST and WFIRST would certainly not have achieved
their top billing in the Astro2010 Decadal Survey if not for their
enormous, broad ranging impact across many fields of astronomy,
including dark energy.
Conversely, the broad astronomical community benefits from the
resources, experties, and effort that the high energy physics
community has brought to these and other projects.

Second, there are important routes to testing dark energy and
modified gravity models that lie outside the scope of the facilities
described here.  These include direct measurements of the Hubble constant,
precision tests of gravity in the solar system or the laboratory,
and studies of gravitational effects in different galactic and
large scale environments (see the Novel Probes Report).
The critical clue to the origin of cosmic acceleration could come
from a surprising direction, including, perhaps, a theoretical
breakthrough.  However, ``mainstream'' dark energy experiments offer
a clear path towards steadily improving precision of cosmic expansion
and structure growth measurements, thus allowing ever more stringent
tests of a wide class of theories of dark energy and modified gravity.

\vfill\eject

\section{The Baryon Oscillation Spectroscopic Survey (BOSS)}

BOSS is the largest of the four surveys that comprise SDSS-III
(SDSS = Sloan Digital Sky Survey).  Its defining goals are to
measure the distance-redshift relation $D_A(z)$ and expansion rate
$H(z)$ with percent-level precision through baryon acoustic oscillations
(BAO), using luminous galaxies to probe redshifts $z \approx 0.2-0.7$
and the \lya\ forest to probe redshifts $z \approx 2 - 3.5$.
BOSS builds on the Luminous Red Galaxy (LRG) survey 
\cite{eisenstein01} of SDSS-I and II \cite{york00}, which enabled
the first detection of BAO (\cite{eisenstein05}, alongside
a contemporaneous detection in the 2dF Galaxy Redshift Survey
by \cite{cole05}) and its refinement into a tool for precision
cosmology.  


To enable BOSS, the SDSS-III Collaboration undertook a major upgrade
(largely funded by DOE) of the original SDSS spectrographs.
The BOSS spectrograph followed the same basic design as SDSS, using aluminum
plates supported in the telescope focal plane by aluminum cast cartridges.
As with SDSS, plates are plugged manually with optical fibers that feed
the two spectrographs mounted to the back of the primary mirror.
The BOSS spectrograph was built with new optical elements and better CCDs to increase throughput
and 50\% more fibers (increased from 640 to 1000 per plate) to increase the overall sample size
\cite{smee13}.  These upgrades allow BOSS to measure galaxy redshifts
at fainter apparent magnitudes than the SDSS spectrographs,
extending the redshift range of luminous galaxies to
$z \approx 0.7$, increasing the galaxy density to three times the SDSS LRG sample
to fully sample structure on the BAO scale, and covering an area of 10,000 deg$^2$.
Forecasts (described in \cite{eisenstein11} and \cite{dawson12}) predict BAO
measurement precision from the full BOSS galaxy survey of 
1.0\% and 1.1\% on $D_A(z)$ for independent redshift bins centered at 
$z = 0.35$ and $z = 0.6$, respectively.  The forecast precision for $H(z)$ 
at the same redshifts is 1.8\% and 1.6\%.  
BAO analysis of the Data Release 9 sample (DR9),
about 1/3 the size of the
final data set, yield 1.9\% precision on the weighted combination
$(D_A^2/H)^{1/3}$ at $z = 0.57$ \cite{anderson12}.
Analyses that use the full shape
of the galaxy power spectrum instead of isolating the BAO peak
can sharpen the measurement precision by a significant factor
and add constraints on neutrino masses and other cosmological parameters, 
at the price of relying more strongly on models of non-linear galaxy bias
(e.g., \cite{sanchez12}).  Redshift-space distortion (RSD) analysis
of the BOSS sample (e.g., \cite{reid12})
has yielded the tightest constraints to date on
the growth rate of structure and on cosmic geometry via the
Alcock-Paczynski (\cite{alcock79}) effect.

BOSS is also pioneering a novel method of measuring BAO at high
redshift using the 3-dimensional clustering of \lya\ forest absorption
towards a dense grid of high-redshift quasars ($2.15 <z< 3.5$).
Analyses of the DR9 data set have yielded BAO measurements at $z \sim 2.5$ 
with precision of 2-3\% \cite{busca12,slosar13}, the first 
BAO measurement by any method at $z > 1$.  Because RSD strongly
enhances line-of-sight clustering 
in the \lya\ forest, the best constrained combination of 
$H(z)$ and $D_A(z)$ is heavily weighted towards $H(z)$, which is
directly sensitive to the total energy density at the
measurement redshift.  Extrapolating to the size of the final
BOSS data set implies an eventual measurement precision of 1.5\%
on a combination that is approximately $D_A^{0.2}/H^{0.8}$.

SDSS-III is being carried out by an international collaboration 
of several hundred scientists at 40+ institutions in the U.S.,
Europe, Asia, and South America.  U.S. funding comes from
the Alfred P.\ Sloan Foundation, the DOE Office of Science,
the NSF, and the participating institutions.

BOSS has stayed on budget and on or ahead of schedule in hardware
development, software development, and observations, and it
remains on target to meet or exceed its original design goals.
The DR10 data set became public in summer 2013, 
and papers analyzing galaxy and \lya\ forest clustering in
this data set are expected in the second half of 2013.
BOSS will complete all of its observations before summer 2014;
the final data release, including 1.3 million new
galaxy redshifts and spectra of more than 160,000 $z > 2$ quasars,
is planned for December 2014.

More information about BOSS can be found in 

\rref
{Dawson}, K.~S., {Schlegel}, D.~J., {Ahn}, C.~P., {Anderson}, S.~F., {Aubourg},
  {\'E}., {Bailey}, S., et~al. 2013. {The Baryon Oscillation
    Spectroscopic Survey of SDSS-III}. AJ 145, 10. \cite{dawson12}

\rref
{Eisenstein}, D.~J., {Weinberg}, D.~H., {Agol}, E., {Aihara}, H., {Allende
  Prieto}, C., {Anderson}, S.~F., et~al., 2011. {SDSS-III: Massive
    Spectroscopic Surveys of the Distant Universe, the Milky Way, and 
    Extra-Solar Planetary Systems}. AJ 142, 72 \cite{eisenstein11}

\rref
{Anderson}, L., {Aubourg}, E., {Bailey}, S., {Bizyaev}, D., {Blanton}, M.,
  {Bolton}, A.~S., et~al., 2012. {The clustering of galaxies in the
    SDSS-III Baryon Oscillation Spectroscopic Survey: baryon acoustic
      oscillations in the Data Release 9 spectroscopic galaxy sample}. 
     MNRAS 427, 3435 \cite{anderson12}

\vskip 0.2truein

\begin{center}
\includegraphics[width=3truein]{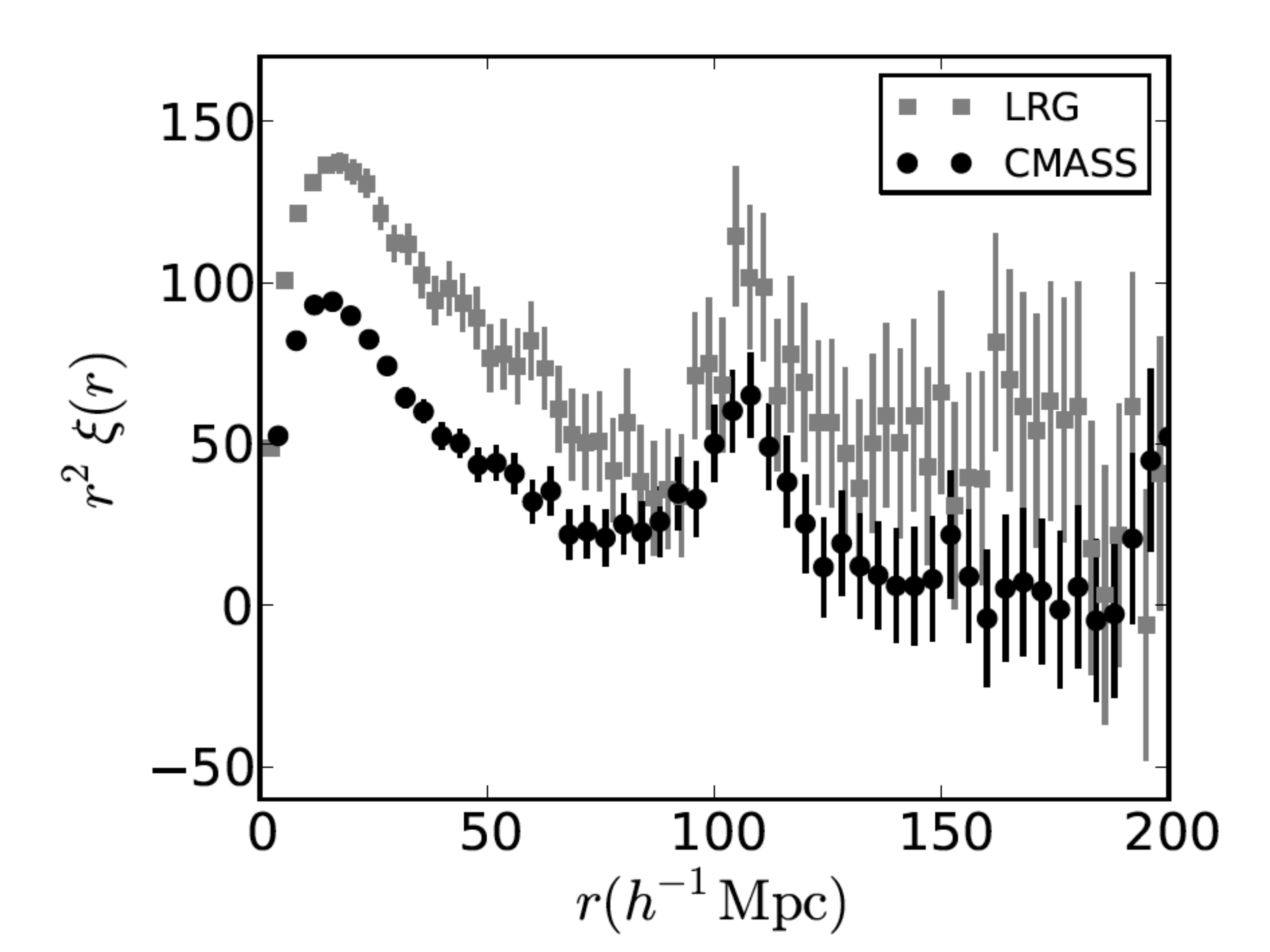}
\hfill
\includegraphics[width=3truein]{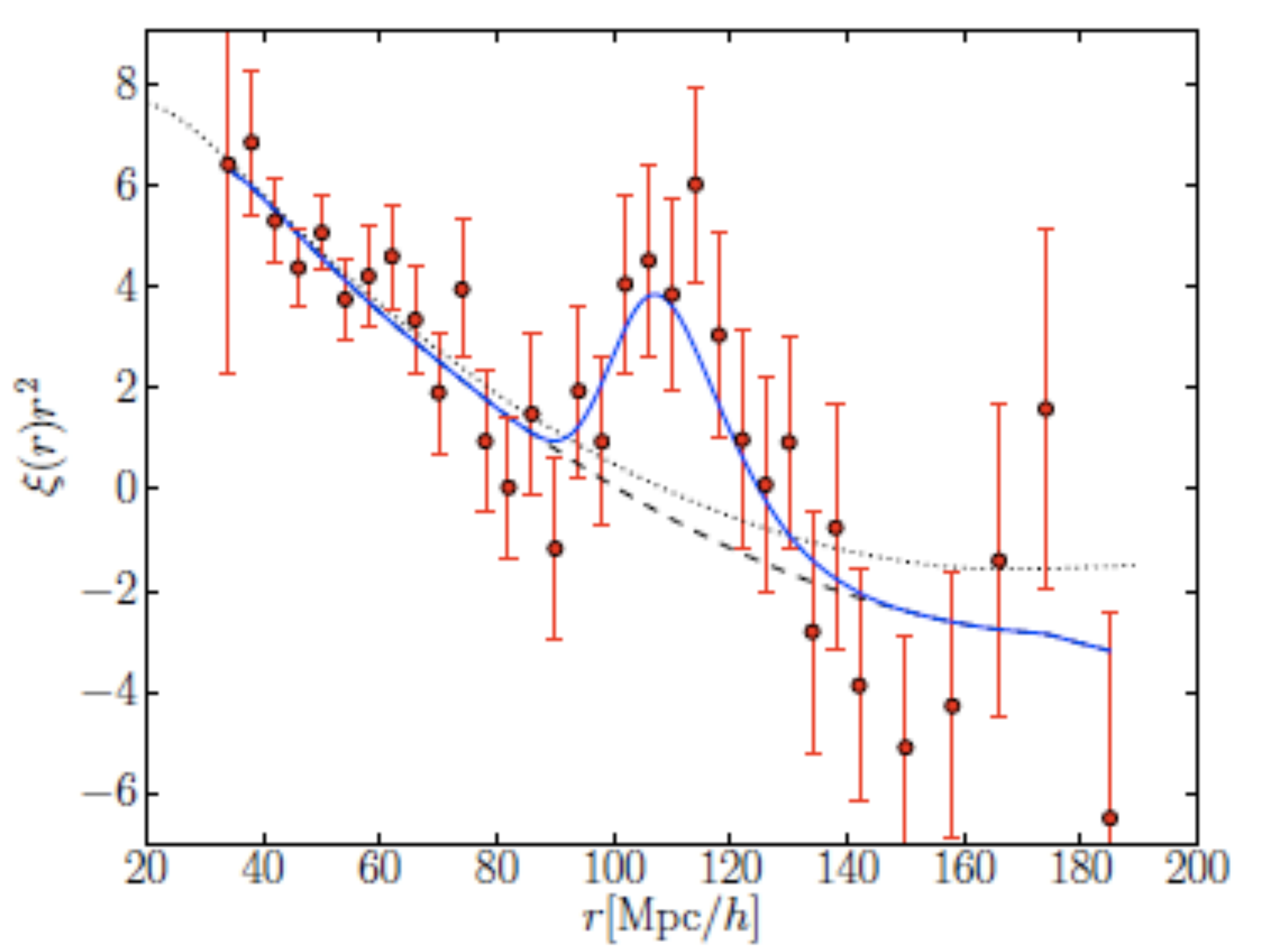}
\end{center}

{\bf Figure 1:}
{\it Left:} The galaxy correlation function, multiplied by $r^2$ to highlight
large scales, from SDSS Luminous Red Galaxies (grey) at $z = 0.35$ 
and BOSS galaxies (black) at $z = 0.57$.  The amplitude of clustering
is different because of the different galaxy types, but the BAO peak
appears at the same comoving scale as expected.  From \cite{anderson12}.
{\it Right:} The correlation function of flux in the \lya\ forest at
$z = 2.5$, showing the BAO peak measured in intergalactic hydrogen absorption.
From \cite{slosar13} (see also \cite{busca12}).

\vfill\eject

\section{The Dark Energy Survey (DES)}

The Dark Energy Survey (DES) is using a powerful new instrument, the
prime focus Dark Energy Camera on the Blanco 4-meter telescope at
Cerro Tololo Inter-American Observatory (CTIO),
to probe the origin of cosmic acceleration using
Type Ia supernovae (SN), weak lensing (WL), galaxy clusters,
and baryon acoustic oscillations (BAO) in photometric galaxy clustering.
The 570-Megapixel DECam, equipped with 74 red-sensitive CCDs, 
has a 3 deg$^2$ field of view.
The camera was commissioned in Sept.-Oct. 2012 
following significant upgrades to the telescope and its control systems.
Science verification observations took place in late 2012-early 2013, and the 
camera is 
performing to specifications.  The survey proper began in August 2013
and will use 525 nights of observation over the ensuing
five years.

DES will image 5000 deg$^2$ of the southern and equatorial sky
in five optical passbands ($grizY$), reaching a depth of approximately
24th magnitude (AB) in each band.  
The imaging depth is about 2 magnitudes deeper than the SDSS,
and the typical image quality is much better ($0.9"$ seeing vs. $1.4"$).
The imaging survey will detect
300 million galaxies, with approximately 200 million WL
shape measurements ($n_{\rm eff} \sim 11\,$arcmin$^{-2})$.
This represents a roughly 40-fold improvement over the CFHTLens
WL survey \cite{heymans13}, which is the largest that exists today,
enabling far more powerful constraints on the growth of dark matter
clustering and the cosmic expansion history.
DES will cover the entire 2500 deg$^2$ footprint of 
the South Pole Telescope Sunyaev-Zel'dovich cluster survey,
and it will detect clusters as enhancements of optical galaxy
counts over its entire footprint, with $\sim 10^5$ clusters 
of mass $M \geq 10^{14} M_\odot$ expected over the full survey
area.  Even more crucial for cluster-based studies of dark energy,
DES will enable stacked
weak lensing measurements (e.g., \cite{sheldon09})
of the virial masses and extended mass profiles of clusters
detected in optical, SZ, or X-ray catalogs, sharply reducing
the mass-calibration uncertainties that are the limiting
systematic of this technique.
Large scale clustering of galaxies in bins of photometric redshift
will enable BAO measurements of the angular diameter distance
$D_A(z)$ out to $z \sim 1.4$, as well as constraints on neutrino
masses and primordial non-Gaussianity.

The DES supernova search will cover 30 deg$^2$ in $griz$, 
returning to each field on a $\sim 5$-day cadence throughout each DES 
observing season.  Forecasts based on detailed simulations
\cite{bernstein11} indicate that DES will discover and measure
high-quality light curves for approximately 4000 Type Ia SNe
over the redshift range $0.5 < z < 1.2$, roughly a factor of ten more than the Supernova Legacy Survey (SNLS).  About 20\% of these
will be followed up with real-time spectroscopy from other
telescopes, while spectroscopic host galaxy redshifts will be
obtained for a larger fraction of the SNe on a longer timescale.

The statistical precision of DES SN, WL, cluster, and photometric
galaxy clustering measurements will far exceed those of any imaging
surveys that exist today.  
Forecasts for dark energy observables depend on assumptions 
about the systematic errors that can be achieved with DES data.
The SN survey has very small statistical uncertainties over nearly
its full redshift range, so it will likely be limited by systematic
uncertainties in photometric calibration, reddening and extinction
corrections, and evolution of the supernova population.
These are already the limiting factors in current SN studies
(e.g., \cite{sullivan11}), but the much larger numbers 
in DES will allow selection of and cross-checks among multiple subsets of 
SNe defined by light curve shape, color, and host galaxy properties,
thus enabling tighter control of systematics. Photometric calibration errors will be controlled in part by repeated measurements of the system throughput and through real-time monitoring of precipitable water vapor at the site. 
For BAO one can forecast the errors on $D_A(z)$ in bins
of photometric redshift, finding expected precision of 1.5-2.5\% 
\cite{blake05}.
SN and BAO measurements at the same redshift have complementary
information content because the SN distance scale is calibrated
in the Hubble flow while the BAO distance scale is calibrated
in absolute units based on cosmological parameters determined
by CMB data --- in effect, SN distances are in $h^{-1}$ Mpc and
BAO distances in Mpc, where 
$h \equiv H_0/100\,{\rm km}\,{\rm s}^{-1}\,{\rm Mpc}^{-1}$ is
the Hubble parameter.

Cluster and WL analyses are sensitive to both geometry and structure
growth, but one can characterize their sensitivity by considering
a constrained model in which the overall amplitude of matter clustering is 
the only free parameter.  Scaling from forecasts in 
\cite{weinberg13} implies that $M \geq 10^{14} M_\odot$ 
clusters calibrated by stacked 
weak lensing in DES can constrain the matter clustering
amplitude (proportional to $\sigma_8$) with 1-2\% precision 
and approximately independent errors in
$\Delta z = 0.1$ redshift bins from $z \approx 0.2$ to $z \approx 1$,
with an aggregate precision better than 0.5\%
(see their figure 30).
Cosmic shear measurements with $N_{\eff} = 2\times 10^8$ could
in principle achieve aggregate precision of 0.2\% or better
on the matter clustering amplitude if limited purely by galaxy
shape noise, but marginalizing over uncertainties in shear calibration,
photometric redshifts, and intrinsic alignments may lead to larger
errors.  

DES is being carried out by an international collaboration 
of more than 200 scientists,
with member institutions in the U.S., the U.K., Spain, Brazil, Germany, and Switzerland.
U.S. funding is being provided by the DOE Office of Science,
the NSF, and the participating institutions.

More information about DES can be found in 

\rref
{\tt https://www.darkenergysurvey.org/reports/proposal-standalone.pdf}

\rref
Bernstein, J.~P., 
Kessler, R., 
Kuhlmann, S.,
Biswas, R.,
Kovacs, E.,
Aldering, G.
et al.\ 2012,  Supernova Simulations and Strategies for 
the Dark Energy Survey,  ApJ, 753, 152 


\begin{center}
\includegraphics[width=5.5truein]{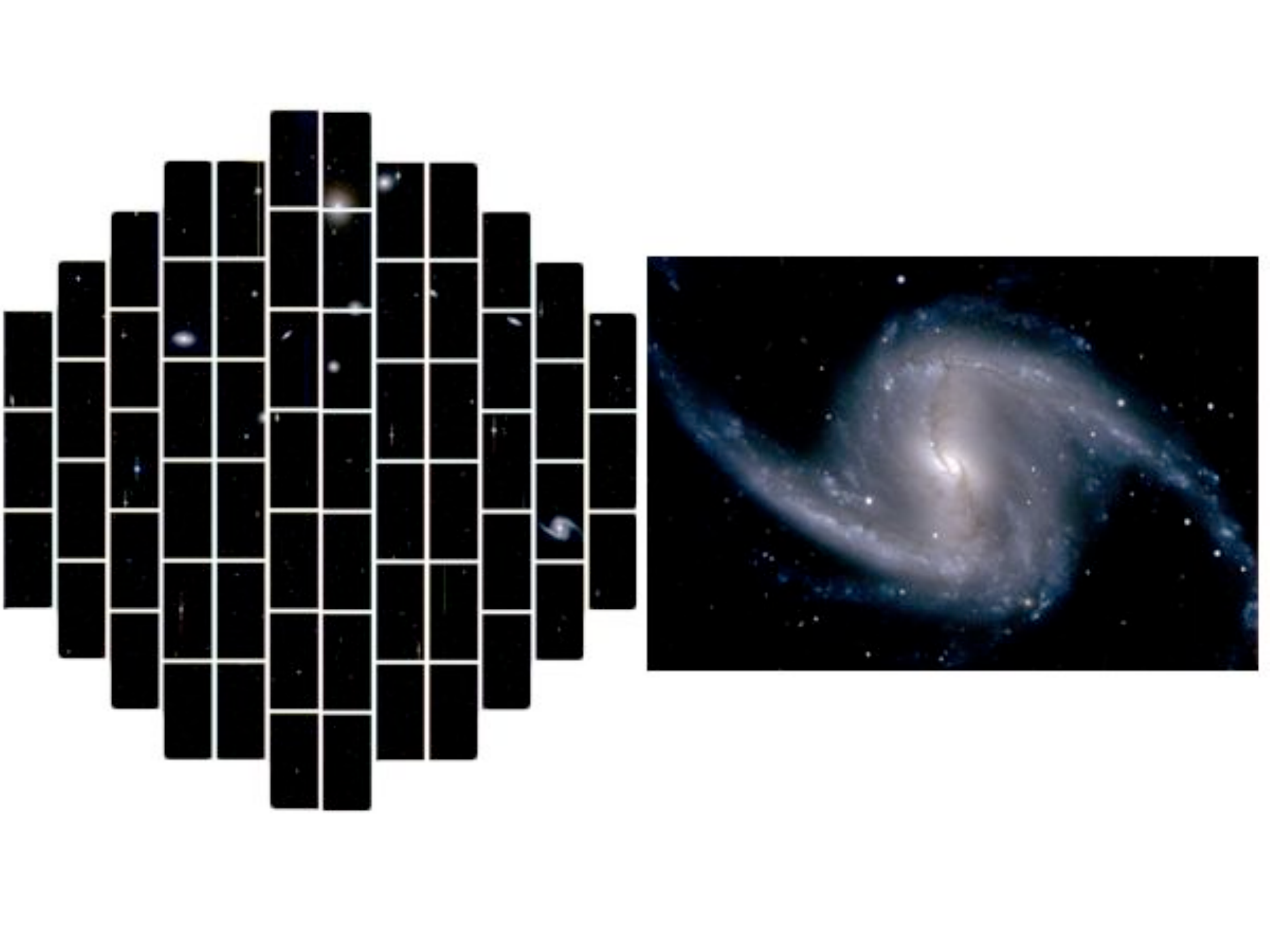}
\end{center}

{\bf Figure 2:}
First light images of the nearby Fornax galaxy cluster
from the 570-Megapixel Dark Energy Camera at CTIO.
The left panel shows the full field of view of the 72-CCD camera.
The right panel zooms in on the image of the galaxy NGC 1365.

\vfill\eject

\section{The Extended Baryon Oscillation Spectroscopic Survey (eBOSS)}

While BOSS will complete its observations in 2014, the Sloan Telescope and
BOSS spectrographs will likely remain the most powerful facility for
wide field spectroscopic surveys until the commissioning of the 
Dark Energy Spectroscopic Instrument (DESI, projected for 2018).
Approved as a major cosmology survey in SDSS-IV (2014--2020), 
eBOSS will capitalize on this premier facility with spectroscopy on a massive
sample of galaxies and quasars in the relatively uncharted redshift
range that lies between the BOSS galaxy sample and the BOSS \lya\ sample.
The targets for eBOSS spectroscopy will consist of:
Luminous Red Galaxies (LRGs: $0.6<z<0.8$) at a density of 50 deg$^{-2}$,
Emission Line Galaxies (ELGs: $0.6<z<1.0$) at a density of 180 deg$^{-2}$,
``clustering'' quasars to directly trace large-scale structure ($1<z<2.2$)
at a density of 90 deg$^{-2}$,
re-observations of faint 
BOSS \lya\ quasars ($2.2<z<3.5$) at a density of 8 deg$^{-2}$,
and new \lya\ quasars ($2.2<z<3.5$) at a density of 12 deg$^{-2}$.

The new target selection for eBOSS has already been proven in a
series of commissioning observations taken with the BOSS spectrograph in
2013.
As shown in those commissioning observations,
all target classes will satisfy classification efficiencies better than 80\%
using the same effective exposure times as BOSS\@.
Quasars and LRGs will be observed 
over 7500 deg$^2$ (5000 deg$^2$ for \lya\ quasars)
and will be the primary primary focus of the program.
The ELG targets will provide a BAO constraint
comparable to the DR9 BOSS galaxy results over a 1500 deg$^2$ area.

Using the techniques for BAO analysis developed in BOSS, eBOSS will measure
both the distance-redshift relation and the evolution of the Hubble parameter
$H(z)$
with these new density tracers, resulting in the first BAO measurements at
$1 < z < 2.2$.
The wide redshift range of the eBOSS tracers will disentangle the
evolution of structure growth from the amplitude of clustering,
thus providing new constraints on GR through RSD analyses.
The expected constraints from BAO and RSD are shown in the table below, where
the fractional distance error from BAO is parametrized by an optimal
combination of $D_A(z)$ and $H(z)$ denoted by a generalized distance
parameter ``R'' and the
growth of structure from RSD is parametrized by $f\sigma_8$,
the product of the matter fluctuation amplitude $\sigma_8$ and
the growth rate $f \equiv d\ln\sigma_8/d\ln a$.

Because the large volume of eBOSS will capture many large-scale modes,
measurements of the matter power spectrum will impose important new
constraints on
non-Gaussian primordial fluctuations in the early universe and on neutrino
masses.
Given Planck results, BAO measurements from BOSS and eBOSS,
5\% $H_0$ priors, and an assumption of zero curvature,
predictions from eBOSS imply a 95\% upper limit $\sum m_{\nu} <0.104$
when combining the
results from the eBOSS LRGs, ELGs and quasars.
The upper limit on neutrino masses derived from eBOSS will be
comparable to the minimum allowed mass in an inverted hierarchy.
Assuming a baseline from Planck measurements and a 5\% $H_0$ constraint,
primordial non-gaussianities of the local form can be constrained
to a precision $\sigma_{fnl}=12$ (68\% confidence) by the combination of
eBOSS LRG, ELG, and quasars.

\begin{table}[htp]
\caption{\small
\label{constrainttable}
Basic parameters expected for each eBOSS sample, together with predictions
for
the effective volumes and fractional constraints on BAO distance measurements
and growth of structure.}
\begin{tabular}{l c c c c c c}
\hline\hline
Sample  & $N_{\rm target}$  & purity & $\bar{n}_{\rm peak}~(h^{-1}{\rm
Mpc})^{-3}$  & V$_{\rm  eff}~Gpc^{3}$& $\sigma_R/R$ &
$\sigma_{f\sigma_8}/f\sigma_8$ \\ \hline
LRGs $0.6<z<0.8$       & 375,000 & 95\% & $1.4 \times 10^{-4}$   &  4.5 &
0.009 & 0.029\\
ELGs $0.6<z<1$          & 270,000 & 80\% & $3.4 \times 10^{-4}$   & 2.1 &
0.018 & 0.035\\
Quasars $1<z<2.2$         & 675,000 & 70\% & $0.21 \times 10^{-4}$   &
4.4 & 0.020 & 0.036\\
BOSS LyA Quasars &   160,000 & --- & --   & -- & 0.015 & --\\
BOSS $+$ eBOSS LyA Quasars &   310,000 & 70\% & --   & -- & 0.011 & --\\
\hline
\end{tabular}
\end{table}

SDSS-IV is being carried out by an international collaboration with
member institutions in the U.S., Europe, Asia, and South America.

More information about SDSS-IV and eBOSS can be found at

\rref
{\tt http://www.sdss3.org/future/sdss4.pdf}

\vskip 0.2truein

\begin{center}
\includegraphics[width=5truein]{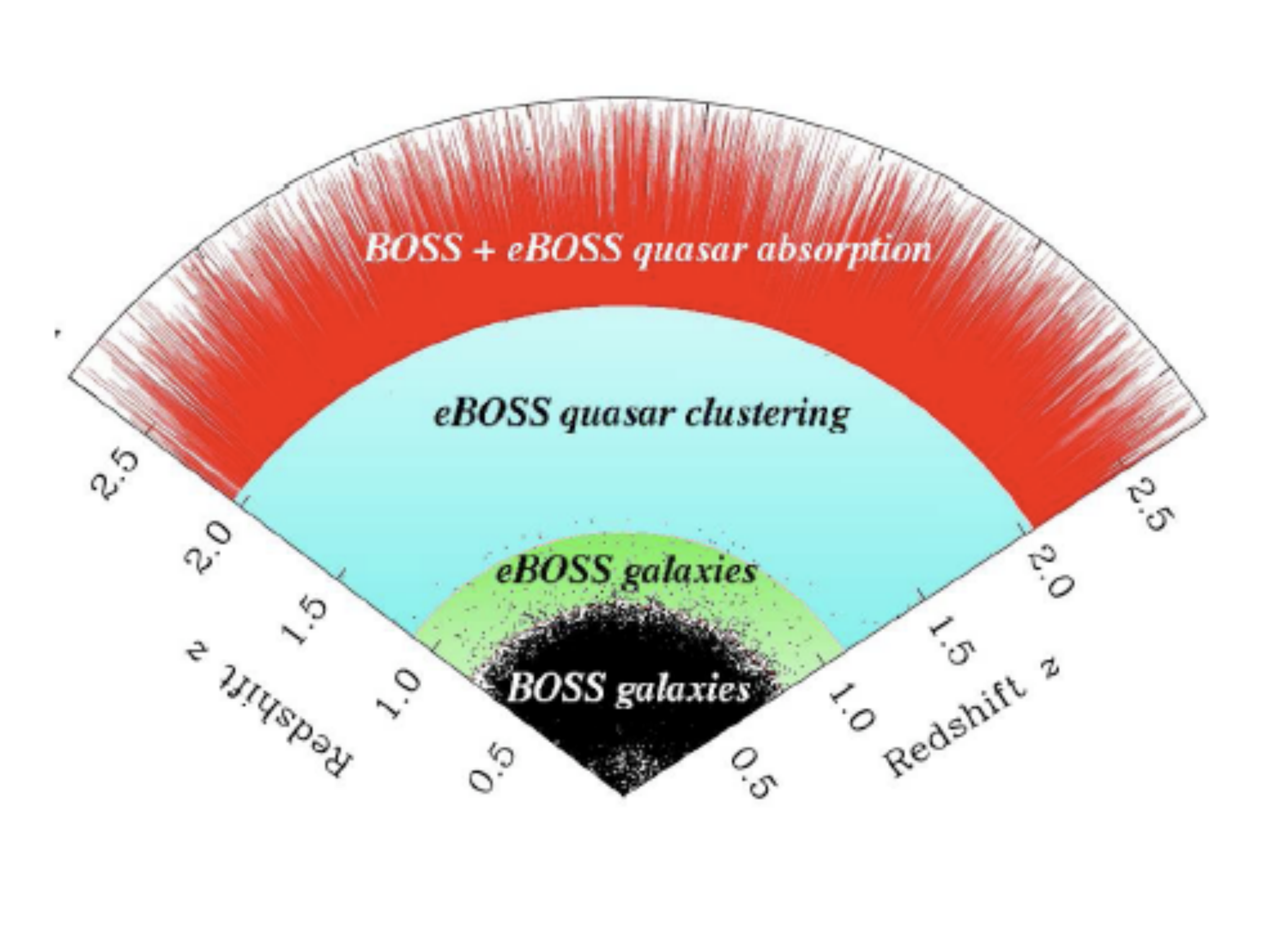}
\end{center}

{\bf Figure 3:}
Complementarity of the eBOSS and BOSS redshift ranges for BAO measurements.
Black dots show a slice through the distribution of BOSS galaxies, with
maximum redshift $z \approx 0.7$.  eBOSS luminous galaxies and emission
line galaxies will probe the redshift range $0.6 < z < 1.1$.
eBOSS quasars will yield the first percent-level BAO measurements
in the redshift range $1 < z < 2$.
At $z > 2$, the addition of eBOSS quasar spectra to BOSS spectra will
sharpen the precision of BAO measurements from the \lya\ forest.

\vfill\eject

\section{The Hobby-Eberly Telescope Dark Energy Experiment (HETDEX)}

The Hobby-Eberly Telescope Dark Energy Experiment (HETDEX) will measure
the expansion rate of the universe in the redshift range
$1.9<z<3.5$, using \lya\ 
emitting galaxies (LAEs). HETDEX consists of a significant upgrade to
the 10-meter Hobby-Eberly Telescope (HET), a new spectroscopic
instrument called VIRUS, and a significant observing campaign. The
upgrade is currently in progress and on schedule for completion
mid-2014. The VIRUS instrument consists of 33,500 optical fibers feeding
75 pairs of spectrographs, providing a wavelength coverage of 350-550nm
with a resolving power of 750. The survey is blind, requiring no
pre-selection of the targets.

The main HETDEX survey will cover 300 square degrees at a
high-declination field with a 20\% fill factor. Thus, about 60 square
degrees of sky will be covered with spectra. The main survey will
produce about 0.7 million LAEs over $1.9<z<3.5$, about 1.0 million [OII]
emitters at $z<0.5$, and significant numbers (many 100,000) of quasars,
stars, and galaxies. The main survey will take three observing seasons
to complete, and thus will run from 2014 to 2017.

HETDEX will also have an equatorial field with a footprint of 150 square
degrees and 30 square degrees of sky with spectra. The equatorial field
will take three observing seasons to complete, and will run from 2014 to
2017. There are plans for extensions of both the high-dec and equatorial
fields in order to increase the full survey volume by a factor of two
over the main survey. The survey extensions will require an additional
two years to complete.

Forecasts indicate that 
the main HETDEX survey will produce a measure of $H(z=2.3)$ to 0.95\% and a
measure of $D_A(z=2.3)$ to 0.95\%. The combined distance measure ($\sigma_R$)
will have an accuracy of 0.85\%. Including both the equatorial and
extension fields will increase the precision on each of these quantities
by about 30\%, but we report the numbers above from the main HETDEX survey as a
conservative estimate. The decision on whether to have
the extension field will be made after the first two years of the HETDEX
survey.

HETDEX will also produce a measure of growth of structure. The main
survey will measure $f\sigma_8$ to an accuracy of 2\% at $z=2.3$. As above,
inclusion of the equatorial and extension fields will improve the
precision by about 30\%.

More information about HETDEX can be found at

{\tt http://hetdex.org}

\vfill\eject

\section{The Subaru Hyper-Suprime Camera and Prime Focus Spectrograph}

The Subaru 8.2-meter telescope at Mauna Kea, Hawaii has a unique prime focus
located 15 m above the primary mirror. The use of this prime focus enables
a wide field of view ($\sim 1.7\,$deg$^2$), exploited first by the Suprime-Cam
instrument. Using this instrument, the high mechanical and optical quality
of the Subaru telescope have been demonstrated. This camera is now being
replaced by a more potent instrument, the Hyper Suprime-Cam (HSC).  This new
camera has been built by National Astronomical Observatory of Japan (NAOJ) and
the Kavli Institute for the Physics and Mathematics of the Universe (KIPMU)
in collaboration with international academic and industrial partners. This
instrument and its 116 CCD camera now being commissioned on the Subaru telescope
will enable a wide and deep optical survey scheduled to start from early
2014. This survey should deliver deep imaging of about 1,500 deg$^2$ in
five bands ($grizy$) with magnitude limits of 26 and 24 in the $i$ and $y$ band
(5$\sigma$, for a point source in 2 arcsec diameter aperture), respectively.

The benefits of devising a multi-object spectrometer sharing the optics
designed for HSC became quickly apparent and led to the Subaru Prime
Focus Spectrograph (PFS) project.  PFS is designed to allow simultaneous
spectroscopy of 2400 astronomical targets over a 1.3 degree hexagonal
field. It shares the Wide Field Corrector and associated Field Rotator and
Hexapod already constructed for the HSC. An array of 2400 optical fibers
is in the Prime Focus Instrument and each fiber tip position is controlled
in-plane by a two-stage piezo-electric Fiber Positioner Cobra system. Each
fiber can be positioned within a particular patrol region such that these
patrol regions fully sample the 1.3 deg field. A Fiber Connector relays light
to four identical fixed-format 3-arm twin-dichroic all-Schmidt Spectrographs
providing continuous wavelength coverage from 380nm to 1.3μm. The blue and
red channels will use two Hamamatsu 2K×4K edge-buttable fully-depleted CCDs
(as in HSC). The near-infrared channel will use a new Teledyne 4RG 4K×4K
HgCdTe 1.7μm cut-off array.

This instrument will be used to observe the same footprint as the HSC
survey. The union of the HSC and PFS survey is known as the Subaru Measurement
of Images and Redshifts (SuMIRe) project. This optical/near-infrared
multi-fiber spectrograph targets will lead to unique cosmological measurements
based on the clustering of galaxies, but also studies of galactic archeology
and of the evolution of galaxies throughout cosmic time. Before and
during the era of extremely large telescopes, PFS will be able to obtain
2400 cosmological/astrophysical targets simultaneously with an 8-10 meter
class telescope. 

The PFS collaboration, led by IPMU, consists of USP/LNA in
Brazil, Caltech/JPL, Princeton, and JHU in USA, LAM in France, ASIAA in Taiwan,
and NAOJ/Subaru.

More information about the Subaru HSC, PFS, and SuMIRE can be found at

\rref
Takada, M., Ellis, R., Chiba, M., et al., arXiv:1206.0737v2

\rref
{\tt http://sumire.ipmu.jp}

\rref
{\tt http://sumire.ipmu.jp/en/2652}

\rref
{\tt http://www.naoj.org/Projects/HSC/index.html}

\vfill\eject

\newcommand\lyaf{LyaF}
\section{The Dark Energy Spectroscopic Instrument (DESI)}

DESI, the Dark Energy Spectroscopic Instrument, will be 
a major leap forward in capability for wide-field spectroscopy
and a powerful experiment for the Cosmic Frontier research program.
DESI is a multi-fiber spectroscopic instrument that will be installed
on the Mayall 4-m telescope to enable massively parallel measurements of
galaxy redshifts.  The resulting 3D map of the Universe will enable baryon
acoustic oscillation (BAO) measurements to chart the expansion history of
the Universe, as well as redshift space distortion measurements to chart
the growth of structure.  DESI fits perfectly within the established Cosmic
Frontiers Dark Energy program:  as a Stage-IV BAO experiment, it complements
the DES and LSST imaging surveys, whose strengths are weak lensing and Type
Ia supernovae, and it fills the hiatus between the end of the DES survey and the
start of LSST.  CD-0 (mission need) was awarded by DOE in September 2012,
CD-1 (conceptual design) is scheduled for January 2014, and first light
is planned in  2018.  LSST and DESI in combination will satisfy the four
Stage IV dark energy probes as established by the Dark Energy Task Force.
The DESI collaboration is currently represented by 160 scientists from 50
institutions in the U.S., Australia, Brazil, China, France, Germany, Korea,
Mexico, Spain, Switzerland and the U.K..

Performing a wide, deep spectroscopic redshift survey of 20 to 30 million
galaxies over $14,000-18,000$ deg$^2$ in a five-year survey requires
a high throughput spectrograph capable of observing thousands of spectra
simultaneously.  The new instrument will be installed on the Mayall Telescope
operated at Kitt Peak, Arizona.  A key DESI innovation is a new prime focus
corrector optic creating an 8 square degree field of view.  Optical fibers
transport the galaxy light from the focal plane to the spectrographs.
The fiber tips are placed with high accuracy using robotic positioners,
and the entire focal plane is reconfigured for each exposure in less than
a minute.  The redshifts of 5000 galaxies are measured with each 20-minute
exposure. The fiber optical cables are split into ten separate bundles
of 500 fibers.  Each bundle then travels 40 meters from the focal plane
to a 500-fiber slit array feeding (ten) separate spectrographs.  The full
spectral bandpass is 360 to 980 nm divided by dichroics between three arms,
attaining a resolution as high as $R = 5200$.

DESI will use a combination of emission-line galaxies (ELGs), luminous red
galaxies (LRGs), and quasars to trace large scale
structure out to redshift $z \approx 1.7$ and \lya\ forest absorption towards
nearly one million high-redshift quasars to measure BAO
at redshifts $1.9 < z < 4$.  As detailed in the Table 
below,
forecast BAO errors on the distance scale are $0.35 - 1.1\%$ per
{\it $\Delta z = 0.2$} redshift bin out to $z = 1.7$, with
an aggregate precision of 0.17\%.  Typical distance errors per bin from the
\lya\ forest, augmented by cross-correlations with quasar density,
are $\sim 1\%$, with an aggregate precision of 0.37\%.
Structure growth is measured via redshift-space distortions in the galaxy and
quasar distributions, with per-bin errors on
the parameter combination $f\sigma_8$ below 2\% over the range $0.2 < z < 1.6$
and aggregate precision of 0.35\%.
The forecasts here use the methods of \cite{seo07} for BAO distance
and \cite{mcdonald09} for RSD, where the RSD uses comoving
wavenumbers with $k < 0.2\,h\,{\rm Mpc}^{-1}$ and bias factors are 
assumed to follow
$b(z)=b_0 \left(D(z=0)/D(z)\right)$, where $b_0=1.7$, 0.84, and 1.2 for
LRGs, ELGs, and QSOs, respectively.

In addition to the measurement of the expansion history of the Universe
and the growth rate of large-scale structure, DESI has an enormously broad
scientific reach.  As one example, DESI will provide an
important constraint on the sum of neutrino masses, one that is complementary
to other approaches that can determine the mass hierarchy or establish
the neutrino mass scale.   Power spectrum
measurements by DESI are sensitive to the sum of neutrino masses.
Because one of the splittings of the squares of the neutrino mass is about
$2.43 \times 10^{-3}{\rm eV}^2$ [KamLAND Collaboration, 2005], we know that
at least one neutrino has a mass of at least 0.05 eV.  If neutrinos have an
inverted mass hierarchy, the minimum sum of the neutrino masses is roughly
twice this (since the other splitting is considerably smaller).
The 1-$\sigma$ limits obtained on the sum of neutrino masses should be
0.024 eV (integrating in k-space up to $k_{\rm max} = 0.1\,h\,{\rm Mpc}^{-1}$).
More stringent constraints (rms of 0.017 eV) are possible if non-linearities
are controlled up to $k_{\rm max} = 0.2\,h\,{\rm Mpc}^{-1}$.

More information about DESI, and the prior proposals BigBOSS and DESpec can
be found in:

{Levi}, M., et~al., 2013,  {The DESI Experiment, a whitepaper for Snowmass
2013},  arXiv:1307.0000.\hfill\break
{Schlegel}, D.,  et~al., 2011,  {The BigBOSS Experiment},  arXiv:1106.1706.
\hfill\break
{Abdalla}, F.,  et~al., 2012,  {The Dark Energy Spectrometer (DESpec)},
arXiv:1209.2451.\hfill\break

\small
\begin{center}
\begin{tabular}{lcccccc}
\hline
\hline
$z$ &
$\frac{\sigma_{R/s}}{R/s}$ &
$V$ &
$\frac{dN_{ELG}}{dz~ d{\rm deg}^2}$ & $\frac{dN_{LRG}}{dz ~d{\rm deg}^2}$ &
$\frac{dN_{QSO}}{dz~ d{\rm deg}^2}$ & $\frac{\sigma_{f\sigma_8}}{f\sigma_8}$\\
 & \%
 &  ${\rm Gpc}^3$ & & & & \%
 \\
\hline
0.1 & 1.77 & 0.81 & 358 &  38 &   5 & 3.26\\
0.3 & 0.82 & 4.72 & 368 & 126 &  22 & 1.60\\
0.5 & 0.56 & 10.33 & 237 & 333 &  31 & 1.34\\
0.7 & 0.41 & 16.08 & 709 & 570 &  34 & 0.94\\
0.9 & 0.36 & 21.17 & 1436 & 442 &  44 & 0.77\\
1.1 & 0.42 & 25.31 & 1393 &  13 &  56 & 0.80\\
1.3 & 0.41 & 28.51 & 1444 &   0 &  69 & 0.81\\
1.5 & 0.51 & 30.87 & 802 &   0 &  81 & 1.04\\
1.7 & 1.10 & 32.52 & 152 &   0 &  80 & 2.36\\
\hline
\hline
\end{tabular}

{\it Constraint forecasts for DESI, courtesy of P.\ McDonald.}
\end{center}

\begin{center}
{
\hfill
\includegraphics[height=2.5truein]{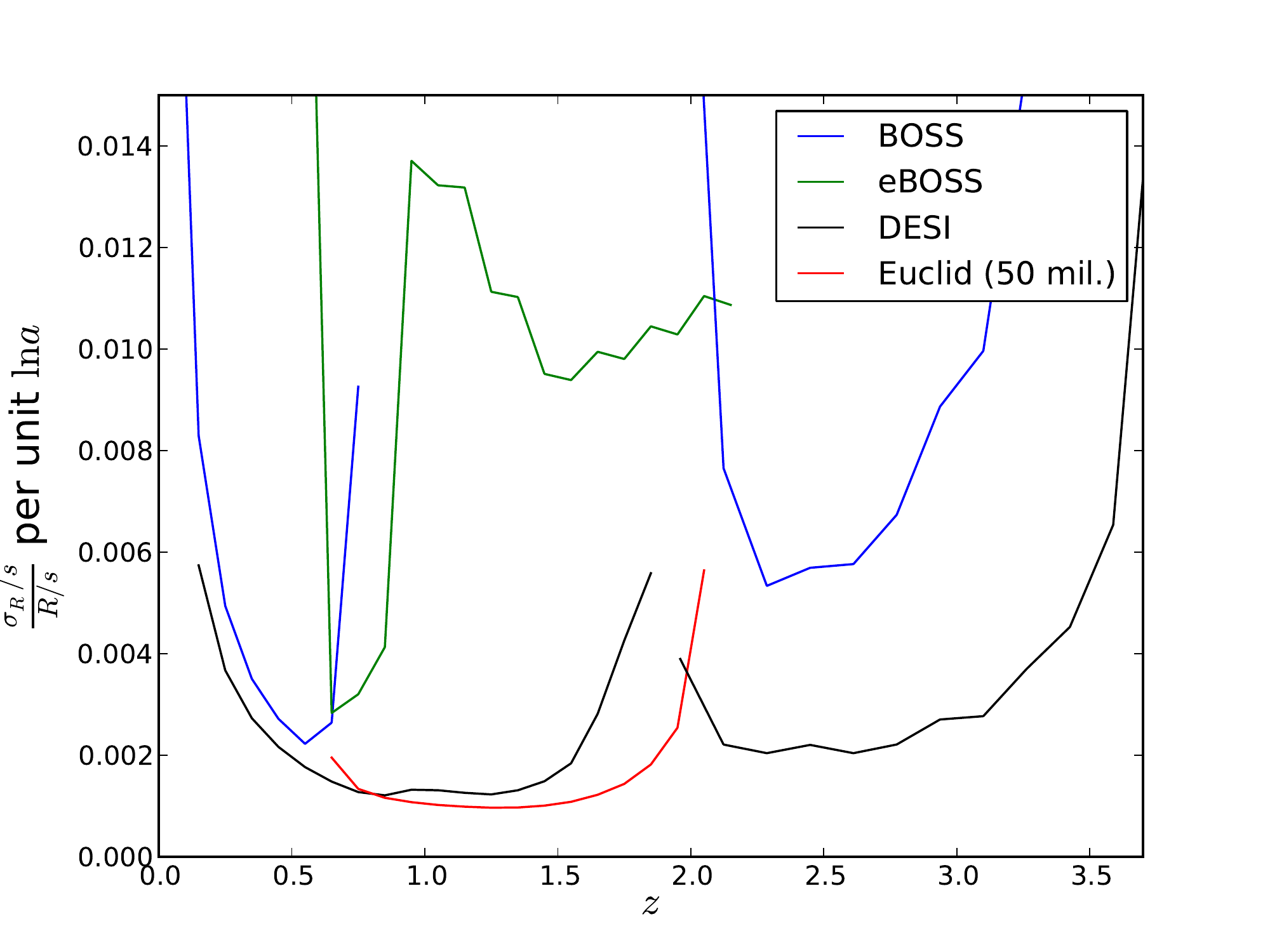}
\hskip 0.1truein
\includegraphics[height=2.5truein]{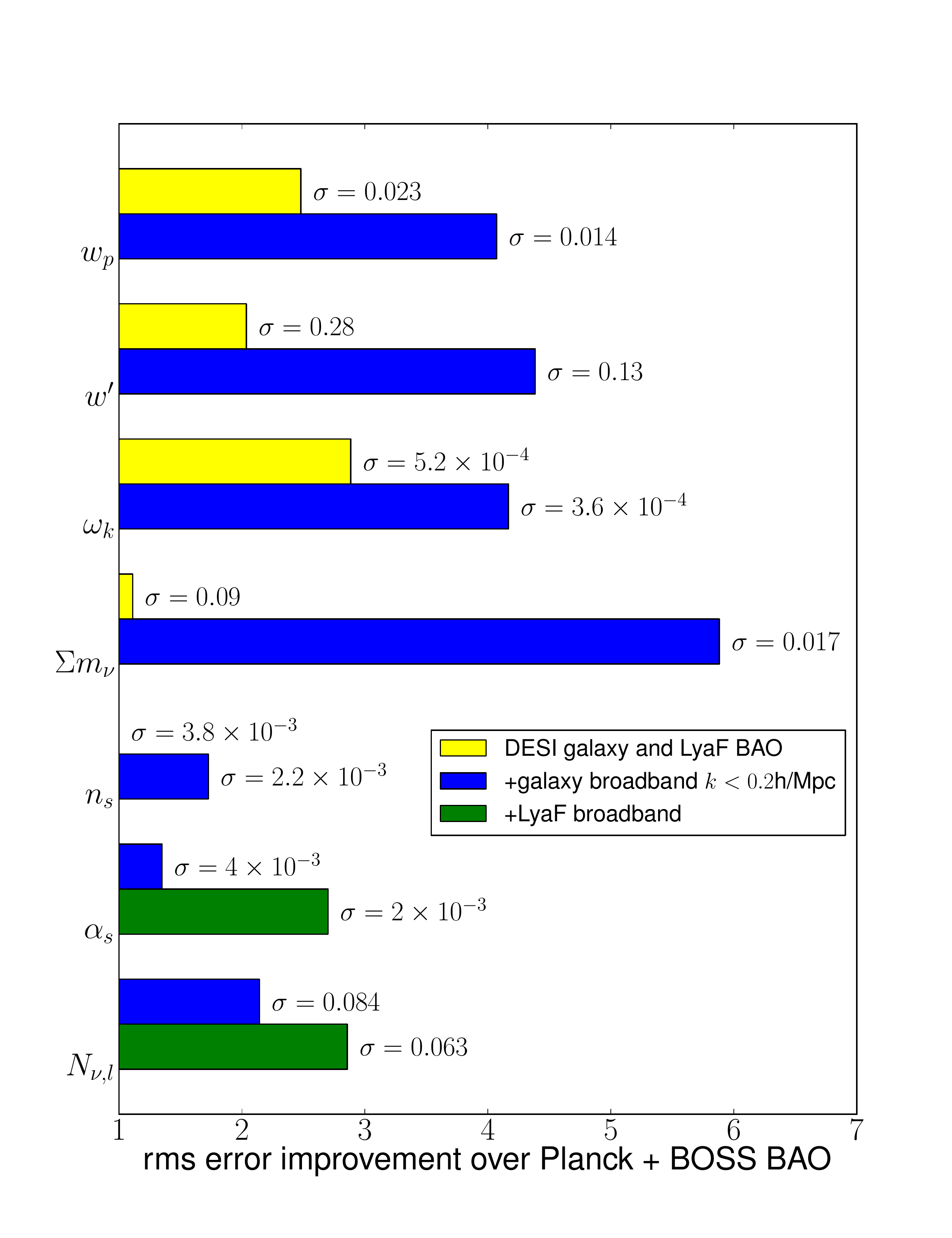}
\hfill
}
\end{center}

{\bf Figure 4:}
{\it Left:} (a) Fractional error on the
BAO distance scale (isotropic dilation factor), as a function of redshift,
{\it per unit $ln(a)$}
(in other words, the effect of any arbitrary redshift bin width $\Delta z$ is
removed in this plot).
Errors from the \lyaf\ measurement, which dominate at $z>1.8$, are
computed following \cite{mcdonald07}, with a modest but significant
additional contribution from cross-correlations with quasar density.
We assume 50 million galaxies for Euclid.
{\it Right:} (b) Improvement in parameters over Planck (final projection,
including polarization) plus BOSS BAO.
Yellow bars show the improvement with DESI BAO,
including \lyaf. Blue
bars show the additional improvement when optimistic broadband power is
used, with $k<0.2\,h\,{\rm Mpc}^{-1}$.
Green additionally adds potential constraints
from broadband (including small-scale 1D) \lyaf\ power.  Dark energy parameters
are defined by the
equation of state $w(z)=w_p + (a_p-a)w^\prime$, where $a_p$ is
chosen to make the errors on $w_p$ and $w^\prime$ independent.
$\omega_k \equiv \Omega_k h^2$ is the equivalent physical density of
curvature. $\sum m_\nu$ is the sum of masses of neutrinos, in eV.
$\alpha_s$ and $n_s$ parameterize the inflationary perturbation
power spectrum, i.e.,  $P_{\rm inflation}(k)
\propto k^{n_s +\frac{1}{2}\alpha_s
\ln\left(k/k_\star\right)}$.

\vfill\eject

\section{Large Synoptic Survey Telescope}

The Large Synoptic Survey Telescope (LSST) is a wide-field, ground-based
telescope, designed to image a substantial fraction of the sky in six optical
bands every few nights.
It is planned to operate for a decade allowing the stacked images to detect
galaxies to redshifts well beyond unity.
The LSST and the survey are designed to meet the requirements
of a broad range of science goals in astronomy, astrophysics and cosmology,
including the study of dark energy.
The LSST was the top-ranked large ground-based initiative in the 2010 National
Academy of Sciences decadal survey in astronomy and astrophysics.

The Dark Energy science goals of LSST are:
\begin{itemize}
\item Weak gravitational lensing: the detection of light from distant sources
due to the bending of space-time by baryonic and dark matter along the line
of sight. Tomographic measurements of weak lensing will provide percent-level
constraints on cosmological parameters.
\item Large-scale structure: the large-scale power spectrum for the
spatial distribution of matter as a function of redshift. This includes the
Baryonic Acoustic Oscillations and the measurement of the distance-redshift
relation. With the enormous number of galaxies detected by LSST, the co-moving
distance will be measured with percent-level precision.
\item Type Ia Supernovae: luminosity distance as a function of redshift
measured with Type Ia SNe as standardizable candles. LSST will discover
and measure at least 500 SNe 1a per season, giving tens of thousands of
well-measured SNe 1a light curves up to z$\sim$1 over the full ten-year survey.
\item Galaxy clusters: the spatial density, distribution, and masses of
galaxy clusters as a function of redshift. LSST will be able to measure the
masses of $\sim$20,000 clusters to a precision of 10\%.
\item Strong gravitational lensing: the angular displacement, morphological
distortion, and time delay for the multiple images of a source object
due to a massive foreground object. LSST will give a sample of $\sim$2600
time-delayed lensing systems, an increase of 100$\times$ compared to the
sample available today.
\end{itemize}

The observatory will be located on Cerro Pachon in northern Chile (near the
Gemini South and SOAR telescopes), with first light expected around 2019.
The survey will yield contiguous overlapping imaging of over half the sky
in six optical bands ($ugrizy$, covering the wavelength range 320-1050 nm).
The LSST camera provides a 3.2 Gigapixel at focal plane array, tiled by 189
4kx4k CCD science sensors with 10$\mu$m pixels.
This pixel count is a direct consequence of sampling the 9.6 deg$^2$
field-of-view
(0.64m diameter) with 0.2x0.2 arcsec$^2$ pixels (Nyquist sampling in the
best expected seeing of $\sim$0.4 arcsec).
The observing strategy for the main survey will be optimized for homogeneity
of depth and number of visits.
The current baseline design will allow about 20,000 deg$^2$ of sky to be
covered using pairs of 15-second exposures in two photometric bands every
three nights on average, with typical 5$\sigma$ depth for point sources of
$r \sim$ 24.5.
The system will yield high image quality as well as superb astrometric and
photometric accuracy for a ground- based survey.
The coadded data within the main survey footprint will have a depth of
$r \sim$ 27.5.
10\% of observing time will be used to obtain improved coverage of parameter
space, such as very deep ($r \sim$ 26) observations taken over the course of
an hour, optimized for detection of faint SNe.

Historically, our understanding of the cosmic frontier has progressed in
step with the size of our astronomical surveys, and in this respect, LSST
promises to be a major advance: its survey coverage will be approximately
ten times greater than that of the Stage III Dark Energy Survey.
The dark energy constraining power of LSST could be several orders of
magnitude greater than that of a Stage III survey \cite{albrecht07}.
LSST will go much further than any of its predecessors in its ability to
measure growth of structure, and will provide a stringent test of theories
of modified-gravity.

While these projections for LSST statistical significance are compelling,
they probably do not capture the true nature of the revolution that LSST
will enable.
The sheer statistical power of the LSST dataset will allow for an all-out
attack on systematics, using a combination of null tests and hundreds of
nuisance parameters and by combining probes.

Beyond tests of systematics, there is a growing sense in the community
that the old, neatly separated categories of dark energy probes will not be
appropriate for next generation surveys.
For example, instead of obtaining constraints on dark energy from cluster
counts and cosmic shear separately, LSST scientists may use clusters and
galaxy-galaxy lensing simultaneously to mitigate the twin systematics of
photometric redshift error and mass calibration~\cite{oguri11}.
A homogeneous and carefully calibrated dataset such as LSST's will be
essential for such joint analyses.

More information on LSST can be found in the following LSST overview papers
and the Science Requirements Document:

\rref
{Ivezi\'{c}}, Z., {Tyson}, J., {Allsman}, R., {Andrew}, J., \& {Angel}, R.,
et~al., 2008. {LSST: from Science Drivers to Reference Design and Anticipated
Data Products}. 0805.2366 \cite{ivezic08}.

\rref
{Abell}, P and the LSST Science Collaborations, 2009, {LSST Science
Book}. 0912.0201 \cite{abell09}.

\rref
{Abate}, A. and the LSST Dark Energy Science Collaboration, 2012, {Large
Synoptic Survey Telescope Dark Energy Science Collaboration}. 1211.0310
\cite{abate12}.

\rref
{Ivezi\'{c}}, Z., \& the LSST Science Collaboration, 2011,
{Large Synoptic Survey Telescope (LSST) Science Requirements Document}.
{\tt http://www.lsst.org/files/docs/SRD.pdf}

\vfill\eject

\section{Euclid}

Euclid is a dark energy satellite mission scheduled for launch in the second
quarter of 2020 \cite{laureijs11}.
Euclid is a European Space Agency
(ESA) Medium Class mission in the Cosmic Vision program.  ESA will provide
the spacecraft, the $1.2$m telescope, and launch on a Russian Soyuz from
Kourou in French Guiana.  The $\sim$ 1200 member Euclid Consortium (EC)
will provide two instruments, a visible imager (VIS) and a near infrared
spectrometer/photmeter (NISP), data processing, and science analysis.  ESA and
the EC will jointly provide an archive to serve the  data to the community.

Euclid is designed for two complementary dark energy probes.  Euclid will
perform a 15,000 deg$^2$ weak gravitational
lensing survey in a single wide optical filter.  The surface density of
resolved galaxies will be between 30 and 40 per square arcminute.  This high
surface density is enabled by the small PSF afforded via space-based, near
diffraction limited observations.  Photometric redshifts for the Euclid
weak lensing galaxies will be measured from  a combination of ground-based
photometry from various sources and 15,000 deg$^2$ near-infrared (NIR)
3 band (1-2 micron) photometric survey taken with Euclid's NISP instrument.
Euclid will also conduct a 15,000 deg$^2$ spectroscopic survey aimed at
galaxy clustering (BAO and RSD). Euclid will use the NISP grism to
measure 50 million redshifts with an accuracy of  $\delta z < 0.001(1+z)$.
In addition to Euclid's wide photometric and spectroscopic survey, it will
perform a 40 square degree deep survey that reaches 2 magnitudes deeper
(in both the optical and NIR) than the wide survey.

The aforementioned surveys, performed over $6.25$ years in the ultra-stable
environment available at the Earth-Sun L2 Lagrange point, will allow Euclid
to address four primary science goals:
\begin{itemize}

\item Measure the equation-of-state of dark energy to high
precision, using both expansion history and structure growth

\item Measure the rate of structure growth to distinguish
General Relativity from modified-gravity theories

\item Test the Cold Dark Matter paradigm for structure formation, and
measure the sum of the neutrino masses to a precision better than $0.02$eV
(when combined with Planck)

\item Greatly improve constraints on cosmological initial conditions
(relative to Planck data alone) to sharpen tests of inflation models
or alternative theories of early-universe physics.

\end{itemize}
Detailed quantitative forecasts for Euclid performance on each of these goals
can be found in \cite{laureijs11}.
Additionally, the Euclid data set of billions of galaxies morphologies
and star formation rates out to $z\sim2$ and tens of millions of redshifts
will form a rich legacy for the study of a wide range of cosmological and
astrophysical phenomena.

Euclid's VIS instrument contains 36 4k$\times$4k e2v CCDs and covers 0.5
square degrees at a pixel scale of $0.1"$ per pixel.  This instrument
has a single, broad RIZ filter.  A dichroic allows simultaneous operation
with the NISP instrument, which can operate in either imaging or grism
spectroscopy mode.  The NISP contains 16 2k$\times$2k Teledyne H2RG detectors.
In imaging mode, there are three filters (Y,J,H) and in slitless spectroscopy
mode there are two grisms and two filters, providing a spectral resolution
of $\frac{\lambda }{\delta \lambda}\sim250$ over 2 pixels.

Euclid data will be released via an archive after a proprietary period.
The current plan, not yet finalized, has three major data releases:
release of the first year of data (about 2500 deg$^2$) 26 months 
after the start of the survey, a release of 7500 deg$^2$ two years
later, and a final release 
containing all of the Euclid data three years after that.

There is already significant involvement in Euclid by US-based
scientists. During the Euclid development phase LBNL (DOE) gave support to the
Euclid Consortium's efforts to define the NISP architecture in exchange for 7
LBNL scientists to join the EC, and LBNL was given a non-voting membership on
the EC's governing Euclid Consortium Board. There were also 7 US scientists
who were granted EC membership due to their involvement in Euclid from
its inception.  In 2013, NASA officially joined Euclid by providing 20
``triplets'' (16 flight and four spares), which consist of a characterized
H2RG detector, a cryogenic cable, and cold readout electronics.  In exchange
for this hardware contribution to NISP, NASA was allowed to select 40 US
investigators to become EC members, and NASA was granted a voting member of
the Euclid Consortium Board and a seat on the 13-member ESA Euclid Science
Team.

More information about Euclid can be found in

\rref
Euclid Definition Study Report, 
Laureijs, R., Amiaux, J., Arduini, S., et al. 2011, 
arXiv:1208.3369
\cite{laureijs11}

\rref
{\tt http://sci.esa.int/science-e/www/area/index.cfm?fareaid=100}

\begin{center}
\hfill
\includegraphics[width=3.1truein]{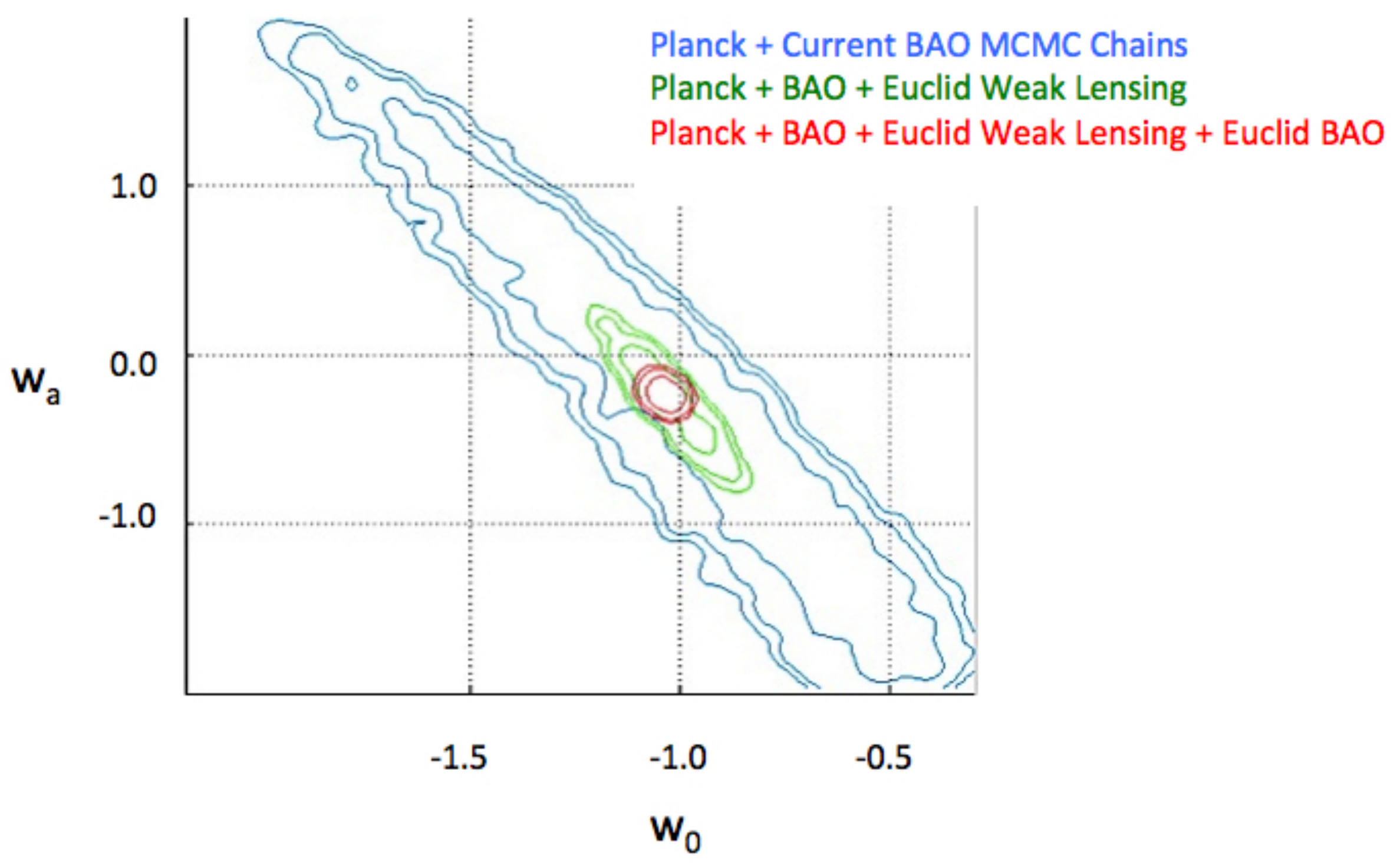}
\hskip 0.1truein
\includegraphics[width=3.1truein]{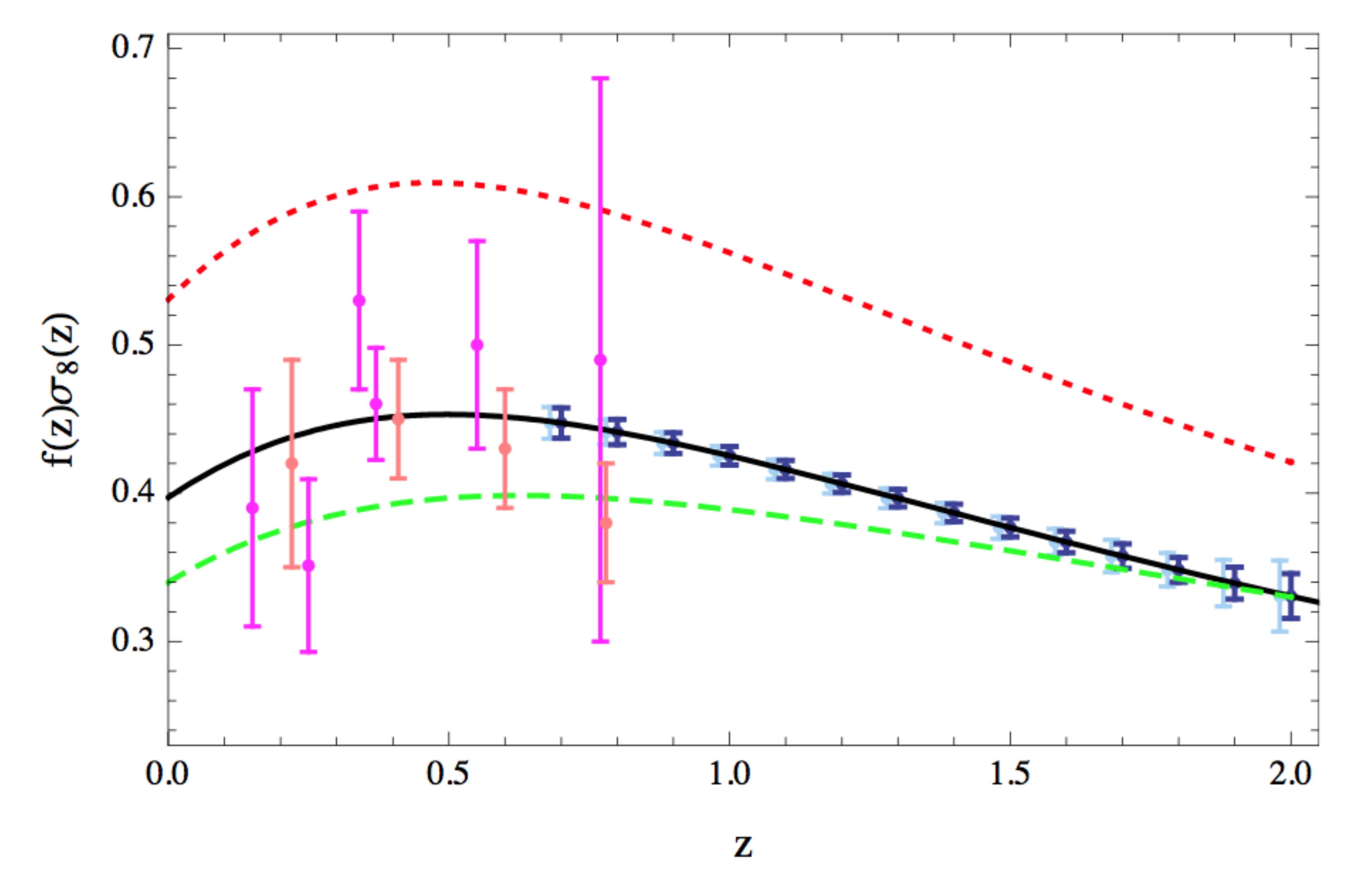}
\hfill
\end{center}

{\bf Figure 5:}
{\it Left:} Projected constraints on parameters of the dark energy
equation of state for combinations of Planck CMB data with current
BAO measurements (blue), adding projected Euclid WL measurements 
(green), and adding projected Euclid WL and BAO measurements (red).
Modeling details are given in \cite{laureijs11}, though this 
figure (courtesy of the Euclid Consortium) uses updated performance
estimates.
{\it Right:} Constraints on the structure growth parameter combination
$f(z)\sigma_8(z)$, 
comparing projections for Euclid redshift-space distortion measurements (blue)
to existing constraints from the WiggleZ survey (orange) and earlier
studies (magenta).  Figure from \cite{majerotto12}.

\vfill\eject

\section{The Wide Field Infrared Survey Telescope (WFIRST)}

The Astro2010 Decadal Survey of astronomy recommended a 
Wide Field Infrared Survey Telescope as its top ranked priority
for a large space mission, recognizing that improvements in infrared
detector technology enabled a mission that would achieve dramatic
advances across a wide range of astrophysics.  Pinning down the
properties of dark energy is one of the primary science goals
of WFIRST, and its design builds heavily on earlier plans for
a DOE-NASA Joint Dark Energy Mission (JDEM), which in turn built
on DOE's proposed SuperNova Acceleration Probe (SNAP).
Following Astro2010, NASA commissioned a Science Definition Team (SDT)
to develop a detailed plan for WFIRST.  The design reference mission
(DRM1) described in the SDT's final report \cite{green12}
uses a 1.3m unobstructed telescope and an infrared camera with
a 0.375 deg$^2$ field of view for imaging and slitless spectroscopy.
More recently, a new SDT assessed an implementation of
WFIRST using one of the Hubble-quality, 2.4m telescopes offered
to NASA after the discontinuation of the national security program 
for which they were originally built.  This implementation, known
as WFIRST-AFTA (or informally as WFIRST-2.4),
appears the most likely to go forward, so it is the
one we describe here.

As described in the SDT report \cite{spergel13}, WFIRST-AFTA would
use a 300-megapixel IR camera with 0.11-arcsec pixels and a 
a 0.281 deg$^2$ field of view.  In the 6-year prime mission, 
1.9 years is devoted to a high-latitude imaging and spectroscopic
survey of 2000 deg$^2$, and 0.5 years to a supernova survey with
three different tiers of area and redshift.  The high-latitude imaging
survey will measure weak lensing shapes of $\approx 500$ million
galaxies in three bands (roughly J, H, and a shortened K) and
weak lensing mass profiles of 40,000 massive ($M > 10^{14} M_\odot$)
clusters.  The spectroscopic survey will measure H$\alpha$ emission-line
redshifts of 20 million galaxies at $z = 1-2$ and 2 million [OIII]
emission-line galaxies at $z = 2-3$.  The supernova survey will
detect, monitor, and spectroscopically confirm 2700 Type Ia supernovae
over the redshift range $z = 0.1-1.7$.  With the assumptions described
by \cite{spergel13}, the forecast aggregate measurement precision from 
these surveys is:
\begin{itemize}

\item SN distance: 0.20\% at $z < 1$, 0.34\% at $z > 1$

\item BAO $D_A(z)$/$H(z)$: 0.40\%/0.72\% at $z = 1-2$, 1.3\%/1.8\% at 
  $z = 2-3$

\item Growth constraints from WL/clusters: 0.16\%/0.14\% at $z < 1$,
  0.54\%/0.28\% at $z > 1$; also 1.2\% from RSD at $z = 1-2$

\end{itemize}
These measurements in turn enable stringent tests of dark energy
and modified gravity models.

As a dark energy mission, WFIRST-AFTA complements Euclid in several ways.
The large aperture and IFU spectrometer on WFIRST-AFTA make it an ideal
facility for supernova cosmology; this is an important probe for WFIRST-AFTA
that is not incorporated in the current Euclid plan.  The WFIRST-AFTA
galaxy redshift survey is ``narrow-deep'' compared to Euclid's
``shallow-wide''; while the total number of galaxies is similar and
the forecast BAO precision comparable, the average space density in the 
WFIRST-AFTA is an order of magnitude higher, enabling techniques that
make more use of small scale clustering or higher order statistics.
The Euclid WL survey is designed for maximum statistical 
power, while the WFIRST-AFTA strategy has a much higher level of 
redundancy for internal cross-checks (three shape measurement bands,
5 - 9 images of each source galaxy).  If cross-checks among WFIRST-AFTA,
Euclid, and LSST show that all of them are achieving statistics-limited
WL measurements, then the combination of the three data sets will be
considerably more powerful than any one in isolation.  WFIRST-AFTA also
provides deep near-IR imaging for photometric redshifts (in combination
with LSST optical), and the galaxy redshift survey and special-purpose
IFU observations may play an important role in photo-$z$ calibration.

While design studies and technology development are ongoing,
the likely formal mission start for WFIRST is FY2017, provided
that the NASA funding wedge opened by the completion of the
{\it James Webb Space Telescope} returns to Astrophysics.
The development schedule is 79 months, including 6 months of reserve,
which would lead to launch in 2023.  There is strong motivation in
the astronomy community to maintain an early WFIRST launch
so that there is overlap between its wide-field survey capability
and JWST's deep follow-up capability.

At the end of its 6-year prime mission, WFIRST-AFTA will remain
an extraordinarily powerful facility for dark energy investigations,
especially if internal tests demonstrate that the SN survey remains
statistics-limited and that single-band WL shape measurements are
adequate.  In this case, $2-4$ years of observations in an extended
mission could improve measurement precision on all axes by a 
factor of two, to confirm any surprising results from earlier studies
and better characterize their physical implications.
WFIRST-AFTA may thus remain an important part of the community's
dark energy portfolio through the 2020s and into the early 2030s.

More information about WFIRST can be found in 

\rref
WFIRST-2.4: What Every Astronomer Should Know.
Spergel, D.,
Gehrels, N.,
Breckinridge, J.,
Donahue, M.,
Dressler, A.,
Gaudi, B. S.,
et al. 2013,
arXiv:1305.5425 \cite{spergel13b}

\rref
Wide-Field InfraRed Survey Telescope-Astrophysics Focused Telescope 
Assets WFIRST-AFTA Final Report.
Spergel, D.,
Gehrels, N.,
Breckinridge, J.,
Donahue, M.,
Dressler, A.,
Gaudi, B. S.,
et al. 2013,
arXiv:1305.5422 \cite{spergel13}

\rref
Wide-Field InfraRed Survey Telescope (WFIRST) Final Report.
Green, J., 
Schechter, P.,
Baltay, C., 
Bean, R.,
Bennett, D.,
Brown R., 
et al. 2012,
arXiv:1208.4012 \cite{green12}


\begin{center}
\includegraphics[width=4.2truein]{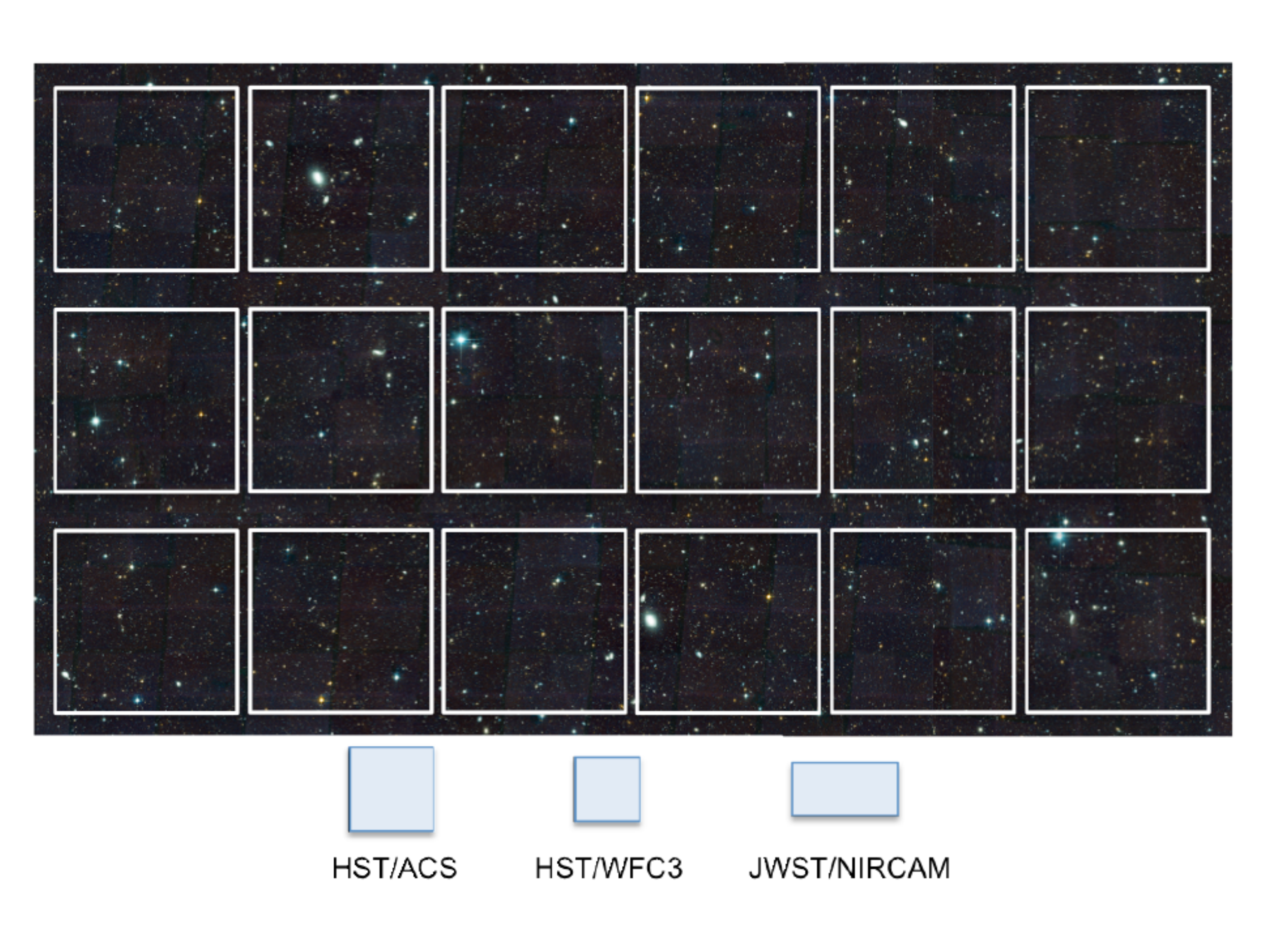}
\end{center}

{\bf Figure 6:}
Field of view of the 300-Megapixel near-IR camera planned for
WFIRST-AFTA.  Each square is a 4k$\times$4k HgCdTe sensor array,
with pixels mapped to 0.11 arcseconds on the sky.  The field of
view extent is 0.79$\times$0.43 degrees.  This is $\sim 200$ times
the area of the IR channel on Hubble Space Telescope's Wide Field
Camera 3, which is shown to scale along with the fields of the
largest imaging cameras on HST (optical) and the James Webb Space
Telescope (IR).

\vfill\eject

\section{Other Opportunities}

The Facilities described in the preceding sections follow the
``main line'' of attack on dark energy outlined in the 2006
Dark Energy Task Force report, from Stage III experiments like
BOSS and DES through to Stage IV experiments like DESI, LSST,
Euclid, and WFIRST-AFTA.  However, this list is by no means an
exhaustive account of the experimental approaches being considered
or, in some cases, actively pursued.  We conclude this document
with a brief overview of other kinds of facilities
or experimental approaches that could play an important role
in dark energy studies over the next 1-3 decades.

{\it Novel Probes of Gravity:} As discussed at length in the
Snowmass white paper on this topic, there are many observational
or laboratory approaches to testing modified gravity theories
in addition to the expansion and growth measurements pursued
by main-line dark energy experiments.  These include precision
tests of the equivalence principle, solar system GR tests,
searches for time or spatial variation of fundamental ``constants,''
and tests of the effective strength of gravity in different
astronomical environments.

{\it Local Calibration and Spectroscopic Follow-Up of Supernova Surveys:}
The large-scale imaging facilities described above are powerful 
tools for discovery and photometric monitoring of supernovae.  
Fully exploiting these surveys requires accurate calibration against
local supernova samples and spectroscopic follow-up of a significant
fraction of the detected supernovae and/or their host galaxies.
The follow-up campaigns have typically been {\it ad hoc} on a
variety of telescopes, but the large size of upcoming imaging
surveys may justify a more systematic approach.  There has
been great progress in constructing well characterized 
local calibrator samples in recent years, but continuing 
attention will be needed to maximize the payoff of future surveys.

{\it Deep Spectroscopic Surveys for Photometric Redshift Calibration:}
Uncertainties in photometric redshift distributions are one of the
most challenging systematics for weak lensing surveys.
Cross-correlation of spectroscopic and imaging surveys
(e.g., eBOSS/DES at Stage III and DESI/LSST at Stage IV)
will be one important approach (see the Snowmass Cross-Correlation
Report), but deep spectroscopic samples from large telescopes
for optimizing algorithms and improving calibrations will
also be crucial.

{\it Radio Surveys for BAO Measurements:} In addition to their optical
and IR emission, galaxies can be detected by the 21cm radio emission
from their neutral hydrogen gas.  Because the BAO scale is large,
radio interferometers can map large scale structure and measure 
BAO without resolving individual galaxies.
This approach has significant technological challenges, particularly
associated with controlling radio frequency interference and foregrounds,
but it is potentially a very efficient way to map high-redshift BAO,
and several experimental efforts are underway or planned.

{\it Balloon-Based Weak Lensing:} Euclid and WFIRST-AFTA will 
mark a major step in weak lensing by moving to the 0.1-arcsecond
resolution that is achievable (over wide fields) only from space.  
A balloon-based experiment could make a first step into this
high angular resolution regime by placing a wide field imaging
platform above most of the atmospheric turbulence that 
affects ground-based images.

{\it Spectroscopic Follow-Up of LSST Imaging:} While the photometric
redshifts from LSST imaging are sufficient for powerful weak lensing
and BAO constraints, the precision on dark energy parameters would
be amplified if moderate resolution spectroscopic redshifts were
available for the bulk of the LSST sample.
Novel instrumentation that might enable relatively inexpensive
ground-based spectroscopy over wide fields, which could
enable such an improvement, is now a subject of
active investigation and development.

{\it Radio Weak Lensing:} A very ambitious radio facility like the
envisioned Square Kilometer Array could make spatially resolved, 
21cm velocity maps for hundreds of millions of galaxies.
These kinematic measurements have the potential to reduce the
shape noise in weak lensing measurements, and in the long run
this approach might yield another leap forward in weak lensing
precision beyond LSST, Euclid, and WFIRST-AFTA.

{\it Standard Sirens:} The gravity wave signals from merging black holes
or neutron stars can be decoded to determine the distance to the source,
providing a ``standard siren'' distance indicator that is rooted in
fundamental physics rather than empirical correlations.
A second-generation (post-LISA) space-based gravity wave interferometer
could achieve the sensitivity and angular resolution needed to 
detect merging compact binaries and unambiguously identify their
galactic hosts over a large fraction of the observable universe.
The limiting source of noise in these distance determinations would
be gravitational lensing by intervening structure, which decreases
at a predictable rate with increasing sample size.
Samples of several hundred thousand sources are achievable in 
principle, which may yield the ultimate in precision measurement
of cosmic acceleration.

\bibliography{DEFacilities/DEFacilities.bib}{}



\end{document}